\documentclass[11pt]{article}
\usepackage{graphicx,amsmath,amssymb,amsxtra,appendix}

\begin{document}

\title{Quark and lepton masses and mixing from a gauged $SU(3)_F$ family symmetry with a light
$\cal{O} ( \text{eV})$  sterile Dirac neutrino}

\author{Albino Hernandez-Galeana\footnote{  On sabbatical leave from:  Departamento de F\'{\i}sica,   ESFM - Instituto Polit\'ecnico Nacional. U. P. "Adolfo L\'opez Mateos". C. P. 07738, Ciudad de M\'exico, M\'exico.  
e-mail: albino@esfm.ipn.mx  }     \\
\small APC laboratory, Universit\'e Paris Diderot, B\^atiment Condorcet, \\
\small 10, rue Alice Domon et L\'eonie Duquet  75205 Paris cedex 13, France. }

\maketitle

\begin{abstract}

In the framework of a complete vector-like and universal gauged $SU(3)_F$ family symmetry,
we report a global region in the parameter space where this approach can account for a realistic
spectrum of  quark masses and mixing in a $4\times 4$ non-unitary $V_{CKM}$,  as well as for the known charged lepton masses and the squared neutrino mass differences reported from neutrino oscillation experiments. The $SU(3)_F$ family symmetry is broken spontaneously in two stages by heavy SM singlet scalars, whose hierarchy of scales yield and approximate $SU(2)_F$ global symmetry associated to the 
almost degenerate boson masses of the order of the lower scale of the $SU(3)_F$ SSB.
The gauge symmetry, the fermion content, and the transformation
of the scalar fields, all together, avoid Yukawa couplings between SM fermions. 
Therefore, in this scenario ordinary heavy fermions, top and bottom quarks and tau lepton, become massive at tree level  from Dirac See-saw mechanisms, while  light fermions, including
light neutrinos, obtain masses from radiative corrections mediated
by the massive gauge bosons of the $SU(3)_F$ family symmetry.
The displayed fit parameter space region solution for fermion masses and mixing yield
the vector-like fermion masses: $M_D \approx 3.2 \,\text{TeV}$, $M_U \approx 6.9 \,\text{TeV}$, 
$M_E \approx 21.6 \,\text{TeV}$, $SU(2)_{F}$ family gauge boson masses
of $\mathcal{O} (2 \,\text{TeV})$, and the squared neutrino mass differences:
$m_2^2-m_1^2 \approx 7.5 \times 10^{-5}\;eV^2$, $m_3^2-m_2^2 \approx 2.2 \times 10^{-3}\;eV^2$, 
$m_4^2-m_1^2 \approx 0.82\;eV^2$.

\end{abstract}




\tableofcontents

\section{ Introduction }

\vspace{3mm}
The origen of the hierarchy of fermion masses and mixing is one of the most important open
problems in particle physics. Any attempt to account for this hierarchy introduce a mass generation mechanism which distinguish among the different Standard Model (SM) quarks and leptons.

\vspace{3mm}
In this report we address the problem of fermion masses and mixing within the framework of 
a broken $SU(3)_F$ gauged family symmetry model \cite{albinosu32004,albinosu3bled} commuting with the SM group\footnote{See \cite{albinosu32004,albinosu3bled} and references therein for some other $SU(3)$ family symmetry model proposals.} . This scenario introduce a hierarchical mass generation mechanism in which light fermions become massive from radiative corrections, mediated by the massive gauge bosons associated to the $SU(3)_F$ family symmetry that is spontaneously broken, while the masses of the top and bottom quarks as well as for the tau lepton, are generated at tree level from "Dirac See-saw"\cite{SU3MKhlopov} mechanisms implemented by the introduction of a new set of $SU(2)_L$ weak singlets $U,D,E$ and $N$ vector-like fermions. Due to the fact that these vector-like quarks do not couple to the $W$ boson, the mixing of $U$ and $D$ vector-like quarks with the SM quarks gives rise to and extended $4\times4$ non-unitary CKM quark mixing matrix \cite{vectorlikepapers}. 

\vspace{3mm}
The extra particles introduced in this scenario include two type of sterile neutrinos: Right handed neutrinos and the vector-like N neutrino.

\section{Model with $SU(3)_F$ flavor symmetry}

\subsection{Fermion content}

\noindent We define the gauge symmetry group

\begin{equation}
G\equiv SU(3)_F \otimes SU(3)_C \otimes SU(2)_L \otimes U(1)_Y
\end{equation}

\noindent  where $SU(3)_F$ is a completely vector-like and universal gauged family symmetry. That is,
the corresponding gauge bosons couple equally to Left and Right Handed ordinary Quarks and Leptons,
including right handed neutrinos. $G_{SM}= SU(3)_C \otimes SU(2)_L \otimes U(1)_Y$ is the "Standard Model" (SM) gauge group, with $g_H$, $g_s$, $g$ and $g^\prime$ the coupling constants. The content of fermions assumes the ordinary quarks and leptons assigned under G as:

\vspace{4mm}

\noindent {\bf Ordinary Fermions:}  $q_{iL}^o=\begin{pmatrix} u_{iL}^o \\ d_{iL}^o \end{pmatrix} \;,\;
l_{iL}^o=\begin{pmatrix} \nu_{iL}^o \\ e_{iL}^o \end{pmatrix}  \; , \; Q = T_{3L} + \frac{1}{2} Y$

\begin{equation*}
\Psi_q^o = ( 3 , 3 , 2 , \frac{1}{3} )_L=\begin{pmatrix} q_{1L}^o \\ q_{2L}^o \\ q_{3L}^o \end{pmatrix}  \quad , \quad \Psi_l^o= ( 3 , 1 , 2 , -1 )_L=\begin{pmatrix} l_{1L}^o \\ l_{2L}^o \\ l_{3L}^o \end{pmatrix} \end{equation*}

\begin{equation*}
\Psi_u^o = ( 3 , 3, 1 , \frac{4}{3} )_R=\begin{pmatrix} u_R^o \\ c_R^o \\ t_R^o \end{pmatrix} \quad ,
\quad \Psi_d^o =(3, 3 , 1 , -\frac{2}{3} )_R=\begin{pmatrix} d_R^o \\ s_R^o \\ b_R^o \end{pmatrix} \end{equation*}

\begin{equation*}
\Psi_e^o = (3 , 1 , 1,-2)_R=\begin{pmatrix} e_R^o \\ \mu_R^o \\ \tau_R^o \end{pmatrix}
\end{equation*}

\noindent where the last entry corresponds to the
hypercharge $Y$, and the electric charge is defined by $Q = T_{3L}
+ \frac{1}{2} Y$. The model also includes two types of extra
fermions:

\noindent {\bf Right Handed Neutrinos:} $ \Psi_{\nu_R}^o = ( 3 , 1 , 1 , 0 )_R= \begin{pmatrix}
 \nu_{e_R} \\  \nu_{\mu_R} \\
\nu_{\tau_R}  \end{pmatrix}  $ ,

\noindent and the $SU(2)_L$ weak singlet vector-like fermions

\vspace{4mm}
\noindent {\bf Sterile Neutrinos: } $\quad N_L^o, N_R^o = ( 1 , 1 , 1 , 0 )  $ ,

\vspace{5mm}
\noindent {\bf The Vector Like quarks:} \\
\begin{equation}
U_L^o, U_R^o = ( 1 , 3 , 1 , \frac{4}{3} ) \quad , \quad
D_L^o, D_R^o = ( 1 , 3 , 1 ,- \frac{2}{3} )  \label{vectorquarks} \end{equation}

\vspace{2mm}
\noindent and
\vspace{2mm}

\noindent {\bf The Vector Like electrons:} $\quad E_L^o, E_R^o = ( 1 , 1 , 1 , -2 ) $

\vspace{3mm}

\noindent The transformation of these vector-like fermions allows the mass invariant mass terms

\begin{equation}
M_U \:\bar{U}_L^o \:U_R^o \,+\, M_D \:\bar{D}_L^o \:D_R^o \,+\, M_E \:\bar{E}_L^o \:E_R^o + h.c. \;,
\end{equation}

and

\begin{equation}
m_D \,\bar{N}_L^o \,N_R^o \,+\, m_L \,\bar{N}_L^o\, (N_L^o)^c \,+\, m_R \,\bar{N}_R^o\, (N_R^o)^c \,+\,  h.c
\end{equation}

\noindent The above fermion content make the model anomaly free. After the definition of the gauge
symmetry group and the assignment of the ordinary fermions in the
usual form under the standard model group and in the
fundamental $3$-representation under the $SU(3)_F$ family symmetry,
the introduction of the right-handed neutrinos is required to
cancel anomalies\cite{T.Yanagida1979}. The $SU(2)_L$ weak singlets
vector-like fermions have been introduced to give masses at tree
level only to the third family of known fermions through Dirac
See-saw mechanisms. These vector like fermions play a crucial role
to implement a hierarchical spectrum for quarks and charged lepton
masses,  together with the radiative corrections.

\section{$SU(3)_F$ family symmetry breaking}

We need to be consistent with low energy Standard Model (SM) and simultaneously we would like to generate and account for the hierarchy of fermion masses and mixing after spontaneously  symmetry breaking (SSB)
down to $SU(3)_C \times U(1)_Q$. Previous basic assumptions of this BSM define the required scalar fields and V.E.V's to achieve the desired symmetry breaking. \\

\noindent  To implement a hierarchical spectrum for charged fermion masses,
and simultaneously to achieve the SSB of $SU(3)_F$, we introduce the
flavon scalar fields: $\eta_i,\;i=2,3$,

\begin{equation*}
\eta_i=(3 , 1 , 1 , 0)=\begin{pmatrix} \eta_{i1}^o\\ \eta_{i2}^o\\ \eta_{i3}^o \end{pmatrix} \;, \quad i=2,3
\end{equation*}

\noindent with  the "Vacuum ExpectationValues" (VEV's):

\begin{equation}
\langle \eta_2 \rangle^T = ( 0 , \Lambda_2 , 0) \quad , \quad
\langle \eta_3 \rangle^T = ( 0 , 0,  \Lambda_3)  \:. \label{veveta2eta3}
\end{equation}

\noindent The above scalar fields and VEV's break completely the
$SU(3)_F$ flavor symmetry. The corresponding $SU(3)_F$ gauge bosons
are defined in Eq.(\ref{SU3lagrangian}) through their couplings to
fermions. Thus, the contribution to the horizontal gauge boson masses
from Eq.(\ref{veveta2eta3}) read

\begin{itemize}

\item $ \langle \eta_2 \rangle :\quad \frac{g_{H_2}^2 \Lambda_2^2}{2} ( Y_1^+ Y_1^- + Y_3^+ Y_3^-) +  \frac{g_{H_2}^2 \Lambda_2^2}{4} ( Z_1^2 + \frac{Z_2^2}{3} - 2 Z_1 \frac{Z_2}{ \sqrt{3}} ) $

\item $ \langle \eta_3 \rangle :\quad \frac{g_{H_3}^2 \Lambda_3^2}{2} ( Y_2^+ Y_2^- + Y_3^+ Y_3^-) + g_{H_3}^2 \Lambda_3^2 \frac{Z_2^2}{3} $
\end{itemize}

\noindent  {\it  These two scalars in the fundamental representation is the minimal set of scalars to break down completely  the $SU(3)_F$ family symmetry}. 

\vspace{5mm}

\begin{tabular}{lcccr}
$SU(3)_F\times G_{SM}$ & $  \xrightarrow{  \langle \eta_3 \rangle \,,\,  \langle \eta_2 \rangle \,,\,  }   $  & $SU(2)_F \;\text{?} \times G_{SM}$     & $ \xrightarrow{   \langle \eta_2 \rangle \,,\, \langle \eta_3 \rangle  }  $ & $G_{SM}$ \\
 &  &  \hspace{-15mm} \text{\bf FCNC \,?}   &     &   \\
 \end{tabular}

\vspace{4mm}
$\Lambda_3 (  \Lambda_2 ) $: 5 very heavy boson masses ($\geq 100\,TeV's$)

\vspace{4mm}
$\Lambda_2 (  \Lambda_3 ) $: 3 heavy boson masses (a few $TeV's$). 

\vspace{4mm}
\noindent Notice that the hierarchy of scales   $ \Lambda_3 ( \Lambda_2 )  \gg \Lambda_2 ( \Lambda_3 )  $ define an "approximate $SU(2)_F$ global symmetry" in the spectrum of $SU(3)_F$ gauge boson masses. To suppress properly the FCNC like, for instance PDG 2016 \cite{PDG2016} : $\mu \rightarrow e \gamma$ ($Br< 5.7\times 10^{-13}$) , $\mu \rightarrow e\,e\,e$  ($Br< 1\times 10^{-12}$) , $K^o - \bar{K^o}$, it is relevant which gauge bosons become massive at the lower scale of the $SU(3)_F$ symmetry breaking.


\vspace{5mm}

\noindent Therefore, neglecting tiny contributions from electroweak symmetry breaking, we obtain the gauge boson mass terms.

\begin{multline}
 M_2^2 \,Y_1^+ Y_1^- + M_3^2 \,Y_2^+ Y_2^- + ( M_2^2 + M_3^2) \,Y_3^+ Y_3^-
+ \frac{1}{2} M_2^2  \,Z_1^2 \\ +\frac{1}{2} \frac{M_2^2 + 4 M_3^2}{3} \,Z_2^2  
- \frac{1}{2}( M_2^2 ) \frac{2}{\sqrt{3}}  \,Z_1 \,Z_2
\end{multline}

\begin{equation} M_2^2= \frac{g_{H}^2 \Lambda_2^2}{2} \quad , \quad M_3^2=\frac{g_{H}^2 \Lambda_3^2}{2}  \quad , \quad y \equiv \frac{M_3}{M_2}= \frac{\Lambda_3}{\Lambda_2} \label{M23}
\end{equation}

\begin{table}[!]
\begin{center} \begin{tabular}{ c | c c }
   &  $Z_1$ & $Z_2$ \\
\hline    \\
$Z_1$ &   $ M_2^2$ &  $ - \frac{  M_2^2}{\sqrt{3}}$ \\
      &           &                            \\
$Z_2$  & $ - \frac{M_2^2}{\sqrt{3}}$  & $\quad \frac{M_2^2+4 M_3^2}{3}$
\end{tabular} \end{center}
\caption{$Z_1 - Z_2$ mixing mass matrix }
\end{table}

\noindent Diagonalization of the $Z_1-Z_2$ squared mass matrix yield the eigenvalues

\begin{eqnarray}
M_-^2=\frac{2}{3} \left( M_2^2 + M_3^2 - \sqrt{ (M_3^2 -  M_2^2)^2 + M_2^2  M_3^2 } \right)_- \label{Mm} \\
\nonumber\\
M_+^2=\frac{2}{3} \left( M_2^2 + M_3^2 +\sqrt{ (M_3^2 -  M_2^2)^2 + M_2^2  M_3^2 } \right)_+ \label{Mp}
\end{eqnarray}

\begin{equation}
M_2^2 \,Y_1^+ Y_1^- +  M_3^2\,Y_2^+ Y_2^- + ( M_2^2 + M_3^2) \,Y_3^+ Y_3^-
+ M_-^2 \,\frac{Z_-^2}{2} +  M_+^2 \,\frac{Z_+^2}{2} \, ,
\end{equation}

\noindent where

\begin{equation}
\begin{pmatrix} Z_1 \\ Z_2  \end{pmatrix} = \begin{pmatrix} \cos\phi & - \sin\phi \\
\sin\phi & \cos\phi  \end{pmatrix} \begin{pmatrix} Z_- \\ Z_+  \end{pmatrix} \label{z1z2mixing}
\end{equation}

\begin{equation*}
 \cos\phi \, \sin\phi=\frac{\sqrt{3}}{4} \,\frac{M_2^2}{\sqrt{ M_2^4 + M_3^2 (M_3^2 -  M_2^2) } }
 \end{equation*}

\vspace{4mm}
\noindent Due to the $Z_1 - Z_2$ mixing, we diagonalize the propagators involving  $Z_1$ and $Z_2$  gauge bosons according to Eq.(\ref{z1z2mixing})

\begin{equation*}
Z_1 = \cos\phi \;Z_- - \sin\phi \;Z_+  \quad , \quad Z_2 = \sin\phi \;Z_- + \cos\phi \;Z_+
\end{equation*}

\section{Electroweak symmetry breaking }

Recently ATLAS\cite{ATLAS} and CMS\cite{CMS} at the Large Hadron Collider announced
the discovery of a Higgs-like particle, whose properties, couplings to fermions
and gauge bosons will determine whether it is the SM Higgs or a member of an extended
Higgs sector associated to a BSM theory.  For electroweak symmetry breaking
we introduction two triplets of $SU(2)_L$ Higgs doublets, namely;

\begin{equation*}
\Phi^u=(3,1,2,-1)=\begin{pmatrix}
\begin{pmatrix} \phi^o\\ \phi^- \end{pmatrix}_1^u \\\\ \begin{pmatrix} \phi^o\\ \phi^- \end{pmatrix}_2^u \\\\
\begin{pmatrix} \phi^o\\ \phi^- \end{pmatrix}_3^u \end{pmatrix} \qquad , \qquad
\Phi^d=(3,1,2,+1)=\begin{pmatrix}
\begin{pmatrix} \phi^+\\ \phi^o \end{pmatrix}_1^d \\\\ \begin{pmatrix} \phi^+\\ \phi^o \end{pmatrix}_2^d \\\\
\begin{pmatrix} \phi^+\\ \phi^o \end{pmatrix}_3^d \end{pmatrix} \, ,
\end{equation*}

\noindent and the VEV?s

\begin{equation*}
 \Phi^u \rangle = \begin{pmatrix}  \langle \Phi_1^u \rangle \\ \langle \Phi_2^u \rangle \\ \langle \Phi_3^u \rangle \end{pmatrix}  \quad , \quad
\langle \Phi^d \rangle= \begin{pmatrix} \langle \Phi_1^d \rangle \\ \langle \Phi_2^d \rangle
\\ \langle \Phi_3^d \rangle \end{pmatrix} \;,
\end{equation*}

\noindent where

\begin{equation*}  \Phi_i^u \rangle = \frac{1}{\sqrt[]{2}}
\begin{pmatrix} v_{ui} \\ 0  \end{pmatrix}  \quad , \quad
\langle \Phi_i^d \rangle = \frac{1}{\sqrt[]{2}}
\begin{pmatrix} 0 \\ v_{di}  \end{pmatrix}  \:.
\end{equation*}

\vspace{2mm}
\noindent The contributions from $\langle \Phi^u \rangle$ and $\langle \Phi^d \rangle$ yield
the $W$ and $Z$ gauge boson masses and mixing with the $SU(3)_F$ gauge bosons

\begin{equation*}
 \frac{g^2 }{4} \,(v_u^2+v_d^2)\,
W^{+} W^{-} + \frac{ (g^2 + {g^\prime}^2) }{8}  \,(v_u^2+v_d^2)\,Z_o^2  
\end{equation*}

\begin{equation*}
\begin{array}{l}
+ \frac{1}{4} \sqrt{g^2 + {g^\prime}^2} \,g_H\,Z_o \,
\left[ \,(v_{1u}^2-v_{2u}^2 -v_{1d}^2+v_{2d}^2)\,Z_1 \right.  \\\\  \left. + (v_{1u}^2+v_{2u}^2 -2v_{3u}^2 -v_{1d}^2-v_{2d}^2+2v_{3d}^2)\,\frac{Z_2}{\sqrt{3}} \right. \\\\
\left. + 2\,(v_{1u} v_{2u}-v_{1d} v_{2d})\,\frac{Y_1^+ + Y_1^-}{\sqrt{2}}  + 2\,(v_{1u} v_{3u}-v_{1d} v_{3d})\,\frac{Y_2^+ + Y_2^-}{\sqrt{2}}  \right.  \\\\  \left.   + 2\,(v_{2u} v_{3u}-v_{2d} v_{3d})\,\frac{Y_3^+ + Y_3^-}{\sqrt{2}}  \right] \\\\
+ \; \text{tiny contributions to the  $SU(3)$ gauge boson masses \,} ,
\end{array}
\end{equation*}

\vspace{4mm}

\noindent $v_u^2=v_{1u}^2+v_{2u}^2+v_{3u}^2$ , $v_d^2= v_{1d}^2+v_{2d}^2+v_{3d}^2$.  Hence,
if we define as usual $M_W=\frac{1}{2} g v$, we may write $ v=\sqrt{v_u^2+v_d^2 } \thickapprox
246$ GeV.

{\small
\begin{equation}
Y_j^1=\frac{Y_j^+ + Y_j^-}{\sqrt{2}}  \quad , \quad  Y_j^\pm=\frac{Y_j^1 \mp i Y_j^2}{\sqrt{2}}
\end{equation}
}

\noindent \emph{ The mixing of $Z_o$ neutral gauge boson with the $SU(3)_F$ gauge bosons modify the couplings of the standard model Z boson with the ordinary quarks and leptons}

\section{ Fermion masses}

\subsection{Dirac See-saw mechanisms}

Now we describe briefly the procedure to get the masses for
fermions. The analysis is presented explicitly for the charged
lepton sector, with a completely analogous procedure for the $u$
and $d$ quarks and Dirac neutrinos. With the fields of particles introduced in
the model, we may write the gauge invariant Yukawa couplings

\begin{equation}
h\:\bar{\psi}_l^o \:\Phi^d \:E_R^o \;+\;
h_2 \:\bar{\psi}_e^o \:\eta_2 \:E_L^o \;+\; h_3 \:\bar{\psi}_e^o
\:\eta_3 \:E_L^o \;+\; M \:\bar{E}_L^o \:E_R^o \;+
h.c \label{DiracYC} \end{equation}

\noindent where $M$ is a free mass parameter ( because its mass
term is gauge invariant) and $h$, $h_2$ and $h_3$ are
Yukawa coupling constants. When the involved scalar fields acquire
VEV's we get, in the gauge basis ${\psi^{o}_{L,R}}^T = ( e^{o} ,
\mu^{o} , \tau^{o}, E^o )_{L,R}$, the mass terms $\bar{\psi}^{o}_L
{\cal{M}}^o \psi^{o}_R + h.c $, where

\begin{equation}
{\cal M}^o = \begin{pmatrix} 0 & 0 & 0 & h \:v_1\\ 0 & 0 & 0 & h \:v_2\\
0 & 0 & 0 & h \:v_3\\
0 & h_2 \Lambda_2  & h_3 \Lambda_3 & M
\end{pmatrix} \equiv \begin{pmatrix} 0 & 0 & 0 & a_1\\ 0 & 0 & 0 & a_2\\
0 & 0 & 0 & a_3\\ 0 & b_2 & b_3 & M
\end{pmatrix} \;. \label{tlmassmatrix} \end{equation}

\noindent Notice that ${\cal{M}}^o$ has the same structure of a
See-saw mass matrix, here for Dirac fermion masses.
So, we call ${\cal{M}}^o$ a {\bf "Dirac See-saw"} mass matrix.
${\cal{M}}^o$ is diagonalized by applying a biunitary
transformation $\psi^{o}_{L,R} = V^{o}_{L,R} \;\chi_{L,R}$. The
orthogonal matrices $V^{o}_L$ and $V^{o}_R$ are obtained
explicitly in the Appendix A. From $V_L^o$ and $V_R^o$, and using
the relationships defined in this Appendix, one computes

\begin{eqnarray}
{V^{o}_L}^T {\cal{M}}^{o} \;V^{o}_R =Diag(0,0,-
\lambda_3,\lambda_4)   \label{tleigenvalues}\\
                                  \nonumber   \\
{V^{o}_L}^T {\cal{M}}^{o} {{\cal{M}}^{o}}^T \;V^{o}_L = {V^{o}_R}^T
{{\cal{M}}^{o}}^T {\cal{M}}^{o} \;V^{o}_R =
Diag(0,0,\lambda_3^2,\lambda_4^2)  \:.\label{tlLReigenvalues}\end{eqnarray}

\noindent where $\lambda_3^2$ and $\lambda_4^2$ are the nonzero
eigenvalues defined in Eqs.(\ref{nonzerotleigenvalues}-\ref{paramtleigenvalues}),
$\lambda_4$ being the fourth heavy fermion mass, and $\lambda_3$ of
the order of the top, bottom and tau mass for u, d and e fermions, respectively.
We see from Eqs.(\ref{tleigenvalues},\ref{tlLReigenvalues}) that at tree level the
See-saw mechanism yields two massless eigenvalues associated to the light fermions:

\section{One loop contribution to fermion masses}

Subsequently, the masses for the light fermions arise through one
loop radiative corrections. After the breakdown of the electroweak
symmetry we can construct the generic one loop mass diagram of
Fig. 1. Internal fermion line in this diagram represent the Dirac see-saw
mechanism implemented by the couplings in Eq.(\ref{DiracYC}). The vertices
read from the $SU(3)_F$ flavor symmetry interaction Lagrangian

\begin{multline} i {\cal{L}}_{int} =   \frac{g_{H}}{\sqrt{2}} \left( \bar{e^{o}} \gamma_{\mu} \mu^{o} Y_1^{+} +
\bar{e^{o}} \gamma_{\mu} \tau^{o} Y_2^{+} + \bar{\mu^{o}} \gamma_{\mu} \tau^{o} Y_3^{+} + h.c. \right)   \\ 
+ \frac{g_{H}}{2} \left( \bar{e^{o}} \gamma_{\mu} e^{o}- \bar{\mu^{o}} \gamma_{\mu} \mu^{o} \right) Z_1^\mu
+  \frac{g_{H}}{2 \sqrt{3}} \left( \bar{e^{o}} \gamma_{\mu} e^{o}+ \bar{\mu^{o}}
\gamma_{\mu} \mu^{o}  - 2 \bar{\tau^{o}} \gamma_{\mu} \tau^{o}  \right) Z_2^\mu  
\label{SU3lagrangian} \end{multline}

\begin{figure}[!]
\centering\includegraphics[width=.7\textwidth]{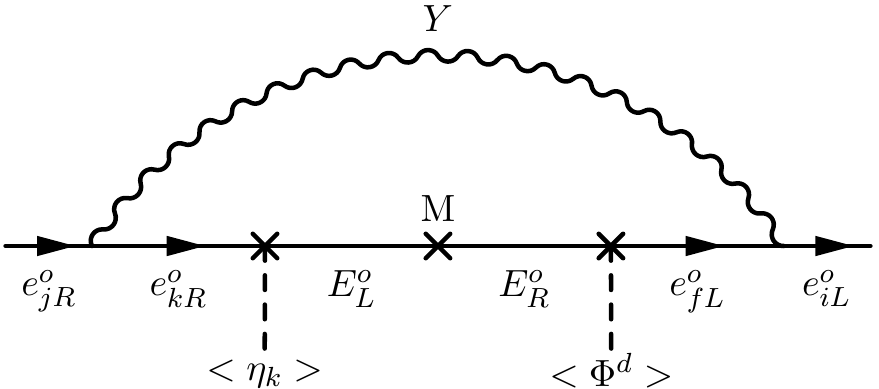}
\caption{ Generic one loop diagram contribution to the mass term
$m_{ij} \:{\bar{e}}_{iL}^o e_{jR}^o$}
\end{figure}

\noindent where $g_H$ is the $SU(3)_F$ coupling constant, $Z_1$, $Z_2$
and $Y_i^j\;,i=1,2,3\;,j=1,2$ are the eight gauge bosons. The
crosses in the internal fermion line mean tree level mixing, and
the mass $M$ generated by the Yukawa couplings in Eq.(\ref{DiracYC})
after the scalar fields get VEV's. The one loop diagram of Fig. 1
gives the generic contribution to the mass term $m_{ij}
\:{\bar{e}}_{iL}^o e_{jR}^o$

\begin{equation} c_Y \frac{\alpha_H}{\pi} \sum_{k=3,4} m_k^o
\:(V_L^o)_{ik}(V_R^o)_{jk} f(M_Y, m_k^o) \qquad , \qquad \alpha_H
\equiv \frac{g_H^2}{4 \pi} \, ,
\end{equation}

\noindent  $M_Y$ is the gauge boson mass, $c_Y$ is a factor
coupling constant, Eq.(\ref{SU3lagrangian}), $m_3^o= - \lambda_3$ and
$m_4^o=\lambda_4$ are the See-saw mass eigenvalues,
Eq.(\ref{tleigenvalues}), and $f(x,y)=\frac{x^2}{x^2-y^2}
\ln{\frac{x^2}{y^2}}$. Using the results of Appendix A, we
compute

\begin{equation} \sum_{k=3,4} m_k^o \:(V_L^o)_{ik}(V_R^o)_{jk} f(M_Y,
m_k^o)= \frac{a_i \:b_j \:M}{\lambda_4^2 - \lambda_3^2}\:F(M_Y) \:,
\end{equation}

\noindent $i=1,2,3$ , $j=2,3$, and $F(M_Y)\equiv
\frac{M_Y^2}{M_Y^2 - \lambda_4^2} \ln{\frac{M_Y^2}{\lambda_4^2}} -
\frac{M_Y^2}{M_Y^2 - \lambda_3^2} \ln{\frac{M_Y^2}{\lambda_3^2}}$. Adding up all
the one loop $SU(3)_F$ gauge boson contributions, we get the mass terms
$\bar{\psi^{o}_L} {\cal{M}}_1^o  \:\psi^{o}_R + h.c.$,

\vspace{1mm}

\begin{equation}
{\cal{M}}_1^o = \left( \begin{array}{ccrc} D_{11} & D_{12} & D_{13}  & 0\\
0 & D_{22} & D_{23} & 0\\ 0 & D_{32} & D_{33} & 0\\
0 & 0 & 0 & 0
\end{array} \right) \:\frac{\alpha_H}{\pi}\; ,
\end{equation}

\vspace{1mm}

\begin{eqnarray*}
D_{11}&=&\mu_{11} (\frac{F_{Z_1}}{4}+\frac{F_{Z_2}}{12} + F_m)+\frac{1}{2} ( \mu_{22} F_1+\mu_{33} F_2 ) \\
D_{12}&=&\mu_{12} (- \frac{F_{Z_1}}{4}+\frac{F_{Z_2}}{12}) \\
D_{13}&=&- \mu_{13} ( \frac{F_{Z_2}}{6}+ F_m ) \\
D_{22}&=&\mu_{22} (\frac{F_{Z_1}}{4}+\frac{F_{Z_2}}{12} - F_m )+\frac{1}{2} ( \mu_{11} F_1+\mu_{33} F_3 ) \\
D_{23}&=&- \mu_{23} ( \frac{F_{Z_2}}{6} - F_m ) \\
D_{32}&=&- \mu_{32} ( \frac{F_{Z_2}}{6} - F_m ) \\
D_{33}&=& \mu_{33} \frac{F_{Z_2}}{3}+\frac{1}{2} ( \mu_{11} F_2 +\mu_{22} F_3 )
\:, \end{eqnarray*}

\vspace{3mm}
\noindent Here,

\begin{equation*}
F_1 \equiv F(M_{Y_1}) \quad,\quad F_2 \equiv F(M_{Y_2}) \quad,\quad F_3 \equiv F(M_{Y_3}) 
\end{equation*}

\begin{equation*}
F_{Z_1}=\cos^2\phi \,F(M_-) + \sin^2\phi \,F(M_+) \; , \; F_{Z_2}=\sin^2\phi \,F(M_-) + \cos^2\phi \,F(M_+)
\end{equation*}

\begin{equation*}
M_{Y_1}^2=M_2^2 \quad,\quad M_{Y_2}^2=M_3^2 \quad,\quad M_{Y_3}^2=M_2^2+M_3^2 \,
\end{equation*}

\begin{equation*}
F_m=\frac{\cos\phi \sin\phi}{2 \sqrt{3}}\, [\, F(M_-)-F(M_+)\,]
\end{equation*}

\noindent with $M_2, M_3, M_-, M_+$ the horizontal boson masses, Eqs.(\ref{M23}-\ref{Mp}),

\begin{equation} \mu_{ij}=\frac{a_i \:b_j \:M}{\lambda_4^2 - \lambda_3^2} = \frac{a_i
\:b_j}{a \:b} \:\lambda_3\:c_{\alpha} \:c_{\beta} \:,\end{equation}

\noindent $c_{\alpha} \equiv \cos\alpha , c_{\beta} \equiv \cos\beta ,
s_{\alpha} \equiv \sin\alpha , s_{\beta} \equiv \sin\beta$, 
Appendix \ref{app} Eq.(\ref{Seesawmixing}). Therefore, up to one loop
corrections we obtain the fermion masses

\begin{equation} \bar{\psi}^{o}_L {\cal{M}}^{o} \:\psi^{o}_R + \bar{\psi^{o}_L}
{\cal{M}}_1^o \:\psi^{o}_R = \bar{\chi_L} \:{\cal{M}}
\:\chi_R \:,\end{equation}

\vspace{1mm} \noindent with ${\cal{M}} \equiv  \left[ Diag(0,0,-
\lambda_3,\lambda_4)+ {V_L^o}^T {\cal{M}}_1^o
\:V_R^o \right]$.

\vspace{5mm}

Using $V_L^o$, $V_R^o$ from Eqs.(\ref{VoL}-\ref{VoR}) we get the mass matrix up to one loop radiative corrections:

\begin{equation} {\cal{M}}= \begin{pmatrix}
m_{11}&m_{12}&c_\beta \:m_{13}&s_\beta \:m_{13} \\
                                                             \\
m_{21}& m_{22} & c_\beta \:m_{23} & s_\beta \:m_{23}\\
                                                             \\
c_\alpha \:m_{31}& c_\alpha \:m_{32} & (-\lambda_3+c_\alpha c_\beta
\:m_{33}) & c_\alpha s_\beta \:m_{33} \\
                                           \\
s_\alpha \:m_{31}& s_\alpha \:m_{32} & s_\alpha c_\beta \:m_{33} &
(\lambda_4+s_\alpha s_\beta \:m_{33})
\end{pmatrix}  \;,\label{massVI}
\end{equation}

\vspace{4mm}
\noindent where

\begin{equation*}
\begin{array}{l}
m_{11} = \delta \,c_1  \,\pi_1    \quad , \quad  m_{21} = -\delta \,s_1 \,s_2  \,\pi_1   \quad , \quad  m_{31} = \delta \,c_2 \,s_1  \,\pi_1   \\\\
m_{12} =\delta \,s_1 \,s_r  \,(c_1 \,c_2 \,c_r \,\Delta + \pi_3)   \\\\
m_{13} =-\delta \,s_1 \, (c_1 \,c_2 \,Fm - c_1 \,c_2 \,s_r^2 \,\Delta + c_r \,\pi_3)  \\\\
m_{22} = \delta (-3 \,c_2 \,c_r \,s_2 \,s_r\,F_m + c_1^2 \,c_2 \,c_r \,s_2 \,s_r \,\Delta + c_2 \,c_r \,\pi_2 + c_1 \,s_2 \,s_r \,\pi_3)   \\\\
m_{23} = \delta (c_2 \,s_2 \,(1 + s_1^2 - 3 s_r^2)\,F_m + c_1^2 \,c_2 \,s_2 \,s_r^2 \,\Delta + c_2 \,s_r \,\pi_2 - c_1 \,c_r \,s_2 \,\pi_3)  \\\\
m_{32} = \delta(-c_r \,s_r\,(-1 + 3 s_2^2) \,F_m + c_r \,(c_2^2 s_1^2 + s_2^2) \,s_r \,\Delta + c_r \,s_2 \,\pi_2
 \vspace{2mm}  \\   \hspace{11mm} - c_1 \,c_2 \,s_r \,\pi_3 )  \\\\
m_{33} = \delta (-  \frac{F_{Z_2}}{6} - (c_2^2 \,s_1^2 - s_2^2 - s_r^2 + 3 s_2^2 \,s_r^2)\,F_m + (c_2^2 \,s_1^2 +
s_2^2) \,s_r^2 \,\Delta  \vspace{2mm}  \\    \hspace{11mm}  + s_2 \,s_r \,\pi_2 + c_1 \,c_2 \,c_r \,\pi_3). 
\end{array}
\end{equation*}

\noindent $s_1,  s_2, s_r ,s_\alpha,  s_\beta , \lambda_3, \lambda_4 $ come from the diagonalization of the tree level mass matrix  ${\cal M}^o$, are defined in Appendix \ref{app}.


 \begin{eqnarray*}
  \delta = \frac{\alpha_H}{\pi} \, c_\alpha \,c_\beta \, \lambda_3 \; ,& \;  \Delta = \frac{1}{4} (F_{Z_2} - F_{Z_1})\, ,& \pi_1 = \frac{1}{2}(c_1 \,c_2 \,c_r \,F_2 + F_1 \,s_2 \,s_r)  
  \end{eqnarray*}

\begin{eqnarray*}
  \pi_2 =  \frac{1}{2}(c_1 \,c_2 \,c_r \,F_3 + F_{Z_1} \,s_2 \,s_r)    \; ,& \; \pi_3 = \frac{1}{2} (c_1 \,c_2 \,c_r \,F_{Z_2} + F_3 \,s_2 \,s_r)
\end{eqnarray*}


\vspace{5mm}



\noindent The diagonalization of ${\cal{M}}$,
Eq.(\ref{massVI})  gives the physical masses for  u and d quarks, e charged leptons and $\nu$ Dirac neutrino
masses. 

Using a new biunitary transformation
$\chi_{L,R}=V_{L,R}^{(1)} \;\Psi_{L,R}$;
\;$\bar{\chi}_L \;{\cal{M}} \;\chi_R= \bar{\Psi}_L \:{V_L^{(1)}}^T
{\cal{M}} \; V_R^{(1)} \:\Psi_R $, with ${\Psi_{L,R}}^T = ( f_1 ,
f_2 , f_3 , F )_{L,R}$ the mass eigenfields, that is

\begin{equation}
{V^{(1)}_L}^T {\cal{M}} \:{\cal M}^T \;V^{(1)}_L =
{V^{(1)}_R}^T {\cal M}^T \:{\cal{M}} \;V^{(1)}_R =
Diag(m_1^2,m_2^2,m_3^2,M_F^2) \:,\end{equation}

\noindent $m_1^2=m_e^2$, $m_2^2=m_\mu^2$, $m_3^2=m_\tau^2$ and
$M_F^2=M_E^2$ for charged leptons. Therefore, the transformation from
massless to mass fermions eigenfields in this scenario reads

\begin{equation} \psi_L^o = V_L^{o} \:V^{(1)}_L \:\Psi_L \qquad \mbox{and}
\qquad \psi_R^o = V_R^{o} \:V^{(1)}_R \:\Psi_R \end{equation}

\vspace{4mm}
\noindent {\bf It  is worth to comment here that neutrinos may also obtain left-handed and right-handed Majorana masses both from tree level and radiative corrections.}

\vspace{5mm}
\subsection{Quark $ V_{CKM} $ and Lepton $U_{PMNS} $  mixing matrices }

Within this $SU(3)_F$ family symmetry model, the transformation from
massless to physical mass fermion eigenfields for quarks and charged leptons is

\begin{equation*} \psi_L^o = V_L^{o} \:V^{(1)}_L \:\Psi_L \qquad \mbox{and}
\qquad \psi_R^o = V_R^{o} \:V^{(1)}_R \:\Psi_R \,,\end{equation*}

\noindent Recall now
that vector like quarks, Eq.(\ref{vectorquarks}), are $SU(2)_L$
weak singlets, and hence, they do not couple to $W$ boson in the
interaction basis. In this way, the interaction of  L-handed up and down quarks; ${f_{uL}^o}^T=(u^o,c^o,t^o)_L$ and
${f_{dL}^o}^T=(d^o,s^o,b^o)_L$, to the $W$ charged gauge boson is

\begin{multline} \frac{g}{\sqrt{2}} \,\bar{f^o}_{u L} \gamma_\mu f_{d L}^o
{W^+}^\mu =  \\  \frac{g}{\sqrt{2}} \,\bar{\Psi}_{u L}\;
[(V_{u L}^o\,V_{u L}^{(1)})_{3\times 4}]^T \;(V_{d L}^o\,V_{d L}^{(1)})_{3\times 4}\;
\gamma_\mu \Psi_{d L} \;{W^+}^\mu \:,
\end{multline}

\noindent $g$ is the $SU(2)_L$ gauge coupling. Hence, the non-unitary $V_{CKM}$ of dimension $4\times
4$ is identified as

\begin{equation} (V_{CKM})_{4\times 4} = [(V_{u L}^o\,V_{u L}^{(1)})_{3\times 4}]^T \;(V_{d L}^o\,V_{d L}^{(1)})_{3\times 4}
\end{equation}

\noindent Similar analysis of the couplings of active L-handed neutrinos and L-handed charged leptons to $W$ boson, leads to the lepton mixing matrix

\begin{equation} ( U_{PMNS} )_{4\times 4}   = [(V_{e L}^o\,V_{e L}^{(1)})_{3\times 4}]^T \;
(U_\nu^o\,U_\nu)_{3\times 4}
\end{equation}

\noindent neglecting neutrino Majorana masses.

\vspace{5mm}

\section{Numerical results} \label{numerical}

\emph{To illustrate the spectrum of masses and mixing from this scenario, let us consider the following
fit of space parameters at the $M_Z$ scale \cite{xingzhang}}

\vspace{3mm}
\noindent Using the input values for the $SU(3)_F$ family symmetry:

\begin{equation} M_2 = 2\,\text{TeV} \quad , \quad M_3 = 2000\,\text{TeV} \quad , \quad \frac{\alpha_H}{\pi}=0.2 \end{equation}

\noindent with $M_2$, $M_3$ horizontal boson masses, Eq.(\ref{M23}), and the coupling
constant, respectively, and the tree level mixing angles

\begin{equation*}
\begin{array}{ll}
s_{1d}=s_{1e}=0.6  \quad & \quad s_{2d}=s_{2e}=0.1047  \\ 
s_{1u}=s_{1 \nu}=0.575341 \quad & \quad s_{2u}=s_{2 \nu}=0.0925127
\end{array}
\end{equation*}

\noindent we obtain the following tree level  ${\cal M}_f^o$, one loop ${\cal M}_{f}$, $f=u,d,e,\nu$ mass matrices, mixing and mass eigenvalues:

\subsection{Quark masses and  $( V_{CKM} )_{4\times 4}$   mixing}

\vspace{5mm}
\noindent {\bf u-quarks:}

\vspace{3mm}
\noindent Tree level see-saw mass matrix:
\begin{equation} {\cal M}_u^o=
\begin{pmatrix}
0	&0	&0	&108921. \\
0	&0	&0	&17589.5 \\
0	&0	&0	&154844. \\
0	&-6.42288 \times 10^6	&462459.	&2.5111 \times 10^6
\end{pmatrix}\,\text{MeV} \,,\end{equation}

\noindent the mass matrix up to one loop corrections:
\begin{equation} {\cal M}_u=
\begin{pmatrix}
7.19764	&-626.533	 &-1479.88	&-3792.15       \\
-0.468392	& -81.7707	&-197.807 &	-506.875 \\
5.04103	&1502.25       &	-172425.   &	12057.2  \\
0.0504129 &15.0233	  & 47.0554     &	6.91226 \times 10^6
\end{pmatrix}\,\text{MeV} \end{equation}

\noindent and the u-quark masses
\begin{equation}
(\,m_u \,,\,m_c \,,\,m_t \,,\,M_U\,)=
(\,  1.396 \,,\,644.835 \;,\,172438\,,\,6.912\times 10^6\,  )\;\text{MeV}
\end{equation}

\vspace{7mm}
\noindent {\bf d-quarks:}

\vspace{3mm}
\vspace{3mm}
\begin{equation} {\cal M}_d^o=
\begin{pmatrix}
0	&0	&0	&2860.87 \\
0	&0	&0	&501.98   \\
0	&0	&0	&3814.49  \\
0	&-2.3645 \times 10^6	&323661.	&2.17117 \times10^6
\end{pmatrix}\;\text{MeV} \end{equation}

\vspace{3mm}
\begin{equation} {\cal M}_d=
\begin{pmatrix}
-4.22954	    &3.26664	   & 26.4239	    &29.045 \\
19.9418	 & -41.6  	      & -57.7027	       & -63.4265 \\
-2.27726	  &-31.0285	     &  -2859.26	          & 755.343 \\
-0.002277	&-0.031028	&0.687179	  &3.2264 \times 10^6
\end{pmatrix}\;\text{MeV} \end{equation}

\vspace{3mm}
\begin{equation}
(\,m_d \,,\,m_s \,,\,m_b \,,\,M_D\,)=
(\, 2.501  \,,\,45.803 \,,\,2860.14   \,,\,3.226 \times 10^6 \,)\;\text{MeV}
\end{equation}

\noindent and the quark mixing
\begin{equation} (V_{CKM})_{4\times 4}=
\begin{pmatrix}
-0.97445	        &0.224576	  & -0.003514	    & -0.000021 \\
-0.224523  	&-0.973562	  &0.042015	    & -0.000010 \\
0.006011	    &0.041720	         & 0.999041         &-0.001233      \\
-0.000219	    &-0.0011268	         &-0.011702	     &   0.000014
\end{pmatrix}  \label{vckm} \end{equation}

\vspace{5mm}
\subsection{Lepton masses and $( U_{PMNS} )_{4\times 4}$ mixing:}

\vspace{3mm}

\vspace{5mm}
\noindent {\bf Charged leptons:}

\begin{equation} {\cal M}_e^o=
\begin{pmatrix}
0	&0	&0	&129165.  \\
0	&0	&0	&22663.9  \\
0	&0	&0	&172220.  \\
0	&-337398.	  &32029.6	&2.16401 \times 10^7
\end{pmatrix}\;\text{MeV} \end{equation}

\vspace{3mm}
\begin{equation} {\cal M}_e=
\begin{pmatrix}
2.2376	   &-66.4545	     & -394.792	    &-6.18239     \\
-0.175708	    &-9.01417	    & -57.8818	     & -0.906422  \\
1.66889	    &164.714	     &  -1693.76	     & 26.556      \\
0.016689  &1.64723	 &16.9589	            & 2.16438 \times 10^7 
\end{pmatrix}\;\text{MeV} \end{equation}

\noindent fit the charged lepton masses:
\begin{equation}
( m_e \,,\, m_\mu \,,\, m_\tau \,,\, M_E ) = ( 0.486 \,,\,102.702\,,\,1746.17\,,\, 2.164\times 10^7\, )\;\text{MeV}
\end{equation}

\vspace{4mm}
\noindent Neglecting for simplicitly the Majorana mass terms, we report the following
Dirac neutrino masses and lepton mixing: 

\vspace{3mm}
\noindent {\bf Dirac neutrino masses:}

\begin{equation} {\cal M}_\nu^o=
\begin{pmatrix}
0	&0	               &0	             & 0.076760 \\
0	&0	               &0	              &  0.012395 \\
0	&0	               &0	             &  0.109124   \\
0	&-0.108392    & -0.264395    &    0.854133
\end{pmatrix}\;\text{eV} \end{equation}

\vspace{3mm}
\begin{equation} {\cal M}_\nu=
\begin{pmatrix}
0.015703	& -0.004190	 & 0.009713	      &0.003179     \\
-0.001021	 &-0.01824	  & -0.005614     & -0.001837  \\
0.010890	&0.004245	&-0.048705	    &-0.002164  \\
0.001539	&0.000600	&-0.000934	      &0.909297
\end{pmatrix}\;\text{eV} \end{equation}

\noindent fit the light neutrino masses:
\begin{equation}
( m_1 \,,\, m_2 \,,\, m_3 \,,\, m_4 ) = ( \, 0.017127  \,,\,      0.0192     \,,\,0.050703\,,\,     0.909309  \, )\, \text{eV}
\end{equation}

\noindent the squared mass differences:

\begin{equation} 
 ( m_2^2 - m_1^2 = \,7.5  \times 10^{-5}  \,,\,m_3^2 - m_2^2 = 2.2  \times 10^{-3}  \,,\,  
 m_4^2 - m_1^2  = 0.826 \,)     \, \text{eV}^2 
\end{equation}

\vspace{4mm}

\noindent and the lepton mixing
\begin{equation} (U_{PMNS})_{4\times 4}=
\begin{pmatrix}
0.610887	      &-0.786302	       & -0.092369	  & 0.003816 \\
-0.709911	     &-0.595411	       & 0.374805	   & 0.032066 \\
-0.349473	      &-0.164968   	& -0.912482	  & -0.133926  \\
0.001821   & 0.000257	& 0.009733	 & 0.001377
\end{pmatrix}  \label{upmns} \end{equation}

\section{Conclusions}

Within the framework of a gauged $SU(3)_{F}$ family symmetry model, we have reported in section 7
a global fit region of the parameter space where this scenario can accommodate a realistic spectrum
for the ordinary quark masses and mixing in a non-unitary $(V_{CKM})_{4 \times 4}$, for the charged lepton masses and the squared neutrino mass differences, within allowed range values reported in PDG 2016 \cite{PDG2016}. Simultaneously, some of extra particles introduced in this scenario; horizontal gauge bosons
and vector-like fermions are predicted to lie within a few TeV's region, and hence, within current LHC energies.

\vspace{3mm}

\noindent \emph{ It is worth to comment that the gauge symmetry $G\equiv SU(3)_F \times G_{SM}$, the fermion content, and the transformation of the scalar fields, all together, avoid Yukawa couplings between SM fermions. So, the scalar fields introduced to break the symmetries in the model: $\eta_2$, $\eta_3$, $\Phi^u$ and $\Phi^d$  couple ordinary fermions with their corresponding  vector-like fermion U, D, E and N,  through the  tree level Yukawa couplings. Therefore, FCNC scalar couplings to ordinary fermions are suppressed
by light-heavy mixing angles, which may be small enough to suppress properly the FCNC
mediated by the scalar fields within this scenario. }

\vspace{5mm}
\section*{Acknowledgements}

It is my pleasure to thank the organizers N.S. Mankoc-Borstnik, H.B. Nielsen, M. Y. Khlopov,
and participants for the stimulating Workshop at Bled, Slovenia. 
The author acknowledge partial support from the "Instituto Polit\'ecnico Nacional",
(Grants from EDI and COFAA) and "Sistema Nacional de Investigadores" (SNI) in Mexico.


\appendix

\section{Diagonalization of the generic Dirac See-saw mass matrix} \label{app}

\begin{equation} {\cal M}^o=
\begin{pmatrix} 0 & 0 & 0 & a_1\\ 0 & 0 & 0 & a_2\\ 0 & 0 & 0 &
a_3\\ 0 & b_2 & b_3 & c \end{pmatrix} \end{equation}

\vspace{1mm} \noindent Using a biunitary transformation
$\psi^{o}_L = V_L^o \:\chi_L$ and  $\psi^{o}_R = V_R^o
\:\chi_R $ to diagonalize ${\cal{M}}^o$, the orthogonal matrices
$V^{o}_L$ and $V^{o}_R$ may be written explicitly as

\begin{equation}
V^{o}_L = \begin{pmatrix}
c_1 &	-s_1\, s_2 &	s_1 \,c_2 \,c_\alpha	& s_1 \,c_2 \,s_\alpha \\
0 & c_2 &	s_2 \,c_\alpha &	s_2 \,s_\alpha \\
-s_1 &	-c_1 \,s_2 &	c_1 \,c_2 \,c_\alpha	& c_1 \,c_2 \,s_\alpha \\
0 &	0 &	-s_\alpha &	c_\alpha
\end{pmatrix}  \label{VoL}
\end{equation}

\begin{equation}
V^{o}_R = \begin{pmatrix}
1 &	0 &	0 &	0  \\
0 & c_r &	s_r \,c_\beta &	s_r \,s_\beta \\
0 &	-s_r &	c_r \,c_\beta	& c_r \,s_\beta \\
0 &	0 &	-s_\beta &	c_\beta
\end{pmatrix}   \label{VoR}
\end{equation}

\begin{equation}
s_1=\frac{a_1}{a_n}  \; , \; c_1=\frac{a_3}{a_n} \; , \;  s_2=\frac{a_2}{a}   \; , \;
c_2=\frac{a_n}{a} \; , \;  s_r=\frac{b_2}{b}   \; , \;  c_r=\frac{b_3}{b}
 \end{equation}

\begin{equation}
a_n=\sqrt{a_1^2+a_3^2} \quad , \quad a=\sqrt{a_1^2+a_2^2+a_3^2} \quad , \quad b=\sqrt{b_2^2+b_3^2}
\end{equation}

\begin{equation}
a_1=a\,s_1\,c_2 \; , \; a_2=a\,s_2 \; , \; a_3=a\,c_1\,c_2 \; , \; b_2=b\,s_r \; , \;
b_3=b\,c_r
\end{equation}

\vspace{3mm}
\begin{equation} \lambda_3^2 = \frac{1}{2} \left( B - \sqrt{B^2 -4D} \right) \quad , \quad  \lambda_4^2 = \frac{1}{2} \left( B + \sqrt{B^2 -4D} \right)
\label{nonzerotleigenvalues}
\end{equation}

\vspace{3mm}
\noindent are the nonzero eigenvalues of
${\cal{M}}^{o} {{\cal{M}}^{o}}^T$ (${{\cal{M}}^{o}}^T
{\cal{M}}^{o}$), and

\begin{eqnarray} B = a^2 + b^2 + c^2 =
\lambda_3^2+\lambda_4^2\quad &, \quad D= a^2
b^2=\lambda_3^2\lambda_4^2 \;,\label{paramtleigenvalues} \end{eqnarray}

\vspace{1mm}

 \begin{eqnarray} \cos{\alpha} =\sqrt{\frac{\lambda_4^2 -
a^2}{\lambda_4^2 - \lambda_3^2}} \qquad , \qquad \sin{\alpha} =
\sqrt{\frac{a^2
- \lambda_3^2}{\lambda_4^2 - \lambda_3^2}} \:,\nonumber \\
                              \label{Seesawmixing}  \\
\cos{\beta} =\sqrt{\frac{\lambda_4^2 - b^2}{\lambda_4^2 - \lambda_3^2}}
\qquad , \qquad \sin{\beta} = \sqrt{\frac{b^2 -
\lambda_3^2}{\lambda_4^2 - \lambda_3^2}} \:.\nonumber \end{eqnarray}


\begin{thebibliography}{99}


\bibitem{albinosu32004} A. Hernandez-Galeana, Rev. Mex. Fis. {\bf Vol. 50(5)},
(2004) 522. hep-ph/0406315.


\bibitem{albinosu3bled} A. Hernandez-Galeana, Bled Workshops in Physics, (ISSN:1580-4992),
{\bf Vol. 16, No. 2}, (2015) Pag. 47; arXiv:1602.08212[hep-ph];
{\bf Vol. 15, No. 2}, (2014) Pag. 93; arXiv:1412.6708[hep-ph];
{\bf Vol. 14, No. 2}, (2013) Pag. 82; arXiv:1312.3403[hep-ph];
{\bf Vol. 13, No. 2}, (2012) Pag. 28; arXiv:1212.4571[hep-ph]; {\bf Vol. 12, No. 2}, (2011) Pag. 41; arXiv:1111.7286[hep-ph]; {\bf Vol. 11, No. 2}, (2010) Pag. 60; arXiv:1012.0224[hep-ph];  Bled Workshops in Physics,{\bf Vol. 10, No. 2}, (2009) Pag. 67; arXiv:0912.4532[hep-ph];

\bibitem{SU3MKhlopov}
Z.G.Berezhiani and M.Yu.Khlopov, {\it Sov.J.Nucl.Phys.} 51 (1990) 739; 935; {\it Sov.J.Nucl.Phys.} 52 (1990) 60; {\it Z.Phys.C- Particles and Fields} 49 (1991) 73;
Z.G.Berezhiani, M.Yu.Khlopov and R.R.Khomeriki, {\it Sov.J.Nucl.Phys.} 52 (1990) 344;
A.S.Sakharov and M.Yu.Khlopov {\it Phys.Atom.Nucl.} 57 (1994) 651;
M.Yu. Khlopov:
\emph{Cosmoparticle physics}, World Scientific, New York
-London-Hong Kong - Singapore, 1999; M.Yu. Khlopov:
\emph{Fundamentals of Cosmoparticle physics}, CISP-Springer, Cambridge, 2011;
Z.G. Berezhiani, J.K. Chkareuli, {\em JETP Lett.} {\bf 35} (612) 1982; {\em JETP Lett.} {\bf 37} (338) 1983; Z.G. Berezhiani, {\em Phys. Lett. B} {\bf 129} (99) 1983.


\bibitem{vectorlikepapers} J.A. Aguilar-Saavedra, R. Benbrik, S. Heinemeyer, and
M. P\'erez-Victoria, arXiv:1306.0572; J.A. Aguilar-Saavedra, arXiv:1306.4432; Jonathan M. Arnold,
Bartosz Fornal and Michael Trott,  JHEP 1008:059, 2010, arXiv:1005.2185 and references
therein.


\bibitem{T.Yanagida1979} T. Yanagida, Phys. Rev. D {\bf 20}, 2986
(1979).

\bibitem{ATLAS} G. Aad \emph{et. al.}, ATLAS Collaboration, Phys. Lett. {\bf B 716}, 1(2012),
arXiv: 1207.7214.

\bibitem{CMS} S. Chatrchyan \emph{et. al.}, CMS Collaboration, Phys. Lett. {\bf B 716}, 30(2012),
arXiv: 1207.7235.

\bibitem{xingzhang} Zhi-zhong Xing, He Zhang and Shun Zhou, Phys. Rev. D {\bf 86}, 013013 (2012).


\bibitem{PDG2016} The Review of Particle Physics (2016),
C. Patrignani et al. (Particle Data Group), Chin. Phys. C, 40, 100001 (2016).


\end{thebibliography}
\end{document}